\documentclass[conference]{IEEEtran}
\IEEEoverridecommandlockouts

\usepackage{cite}

\ifCLASSINFOpdf
  \usepackage[pdftex]{graphicx}
  \DeclareGraphicsExtensions{.pdf,.jpeg,.png}
\else
\fi

\pdfoutput=1

\usepackage{subfigure}
\usepackage{array}
\usepackage{multirow}

\usepackage{amstext}
\usepackage{listings}
\usepackage{color}

\usepackage{algorithm}
\usepackage{algpseudocode}

\begin{document}
%
\title{Evaluating Accumulo Performance for a Scalable Cyber Data Processing Pipeline}

\author{\IEEEauthorblockN{Scott M. Sawyer and B. David O'Gwynn}
\thanks{This work is sponsored by the Assistant Secretary of Defense for Research and Engineering under Air Force contract FA8721-05-C-0002. Opinions, interpretations, conclusions, and recommendations are those of the author and are not necessarily endorsed by the United States Government.}
\IEEEauthorblockA{MIT Lincoln Laboratory\\
Emails: \{scott.sawyer, dogwynn\}@ll.mit.edu}
}

\maketitle

\begin{abstract}

Streaming, big data applications face challenges in creating scalable data flow pipelines, in which multiple data streams must be collected, stored, queried, and analyzed. These data sources are characterized by their volume (in terms of dataset size), velocity (in terms of data rates), and variety (in terms of fields and types). For many applications, distributed NoSQL databases are effective alternatives to traditional relational database management systems. This paper considers a cyber situational awareness system that uses the Apache Accumulo database to provide scalable data warehousing, real-time data ingest, and responsive querying for human users and analytic algorithms. We evaluate Accumulo's ingestion scalability as a function of number of client processes and servers. We also describe a flexible data model with effective techniques for query planning and query batching to deliver responsive results. Query performance is evaluated in terms of latency of the client receiving initial result sets. Accumulo performance is measured on a database of up to 8 nodes using real cyber data.

\end{abstract}


%
\IEEEpeerreviewmaketitle

\section{Introduction}

An increasingly common class of big data systems collect, organize, and analyze data sets in order to gain insights or solve problems in a variety of domains. Interesting data sets are continuously growing, and they must be captured and stored in a scalable manner. 
A big data pipeline is a distributed system that manages the collection, parsing, enrichment, storage, and retrieval of one or more data sources. The fundamental challenge of a big data pipeline is keeping up with data rates while supporting necessary queries for analytics.

To solve this problem, many big data systems use NoSQL distributed databases, such as Apache  Accumulo~\cite{accumulo}, which is designed to scale to thousands of servers and petabytes of data, while retaining the ability to efficiently retrieve entries. Accumulo differs from a traditional relational database management system (RDBMS) due to its flexible schema, relaxed transaction constraints, range-partitioned distribution model, and lack of Structured Query Language (SQL) support~\cite{fuchslecture}. This design trades consistency and query performance for capacity and ingest performance. Compared to other tabular datastores, Accumulo offers proven scalability and unique security features making it attractive for certain applications~\cite{patil2011ycsb}.

This paper considers the Lincoln Laboratory Cyber Situational Awareness (LLCySA) system as a big data pipeline case study~\cite{llcysa2014}. Situational awareness in the cyber domain addresses the problem of providing network operators and analysts a complete picture of activity on their networks in order to detect advanced threats and to protect data from loss or leak. The LLCySA system is a platform for operators and analytics developers that focuses on big data capability and cluster computing. This approach poses several challenges involving the sheer amount of data, the computational requirements of processing the data, and the interactivity of the system.

There is currently considerable related work in big data pipeline implementation and evaluation, as well as related work in the field of cyber situational awareness. Accumulo has been demonstrated to scale to large cluster sizes~\cite{nsa} and to achieve over 100 million inserts per second~\cite{sen2013benchmarking}. 
Query processing is a noted problem in NoSQL databases, with research focusing on query planning~\cite{husain2011heuristics} and providing SQL support~\cite{abouzeid2009hadoopdb}.
In the area of pipeline frameworks, open source projects for distributed stream processing have emerged, such as Apache Storm, which can serve the use case of continuous message processing and database inserts~\cite{storm}. This paper builds on foundational work in the area of data models for schemaless databases\cite{kepner2012dynamic}. Additionally, other network monitoring systems have investigated approaches using NoSQL databases~\cite{sandia}.

In this paper, we describe our application of Accumulo for a cyber big data pipeline, including our data model, storage schema, parallel ingest process, and query processor design. We describe scalability of the ingest process and quantify the impact of our query processing approach on query responsiveness (i.e., the latency to receiving initial query results).
Specifically, we present the following contributions:
\begin{itemize}
\item Characterization of ingest performance in Accumulo as a function of client and server parallelism,
\item Adaptive query batching algorithm that improves responsiveness of large data retrievals, and
\item An approach to query planning in Accumulo with simple, effective optimization guidelines. 
\end{itemize}

This paper is organized as follows. Sec.~\ref{sec:ingest} describes the data model and ingestion process. Next, Sec.~\ref{sec:query} describes our approach for query planning and batching. Sec.~\ref{sec:results} presents quantitative results for ingest and query experiments, and Sec.~\ref{sec:conclusions} discusses conclusions from these results.

\section{Data Model and Ingest}
\label{sec:ingest}

Accumulo is a key--value store in which records comprise a key (actually a tuple of strings, including a \emph{row ID} and \emph{column qualifier}) and a corresponding value. Since keys contain row and column components, the schema is often thought of as a table. Keys are stored in lexicographically sorted order and distributed across multiple database servers. A master server maintains an index of the sorted key splits, which provides a way to identify the location of a given record. Each key--value pair is called an \emph{entry}, and entries are persisted to a distributed filesystem in an indexed sequential access map (ISAM) file, employing a B-tree index, relative key encoding, and block-level compression. A set of contiguous entries comprise a tablet, which is stored on one of the database \emph{tablet servers}. Because the entries are sorted by key at ingest time, retrieving a value given a key can be accomplished very quickly, nearly independently of the total number of records stored in the system.

LLCySA ingests data sources containing information about network events (e.g., authentication, web requests, etc.). The system stores more than 50 event types derived from sources such as web proxy, email, DHCP, intrusion detection systems, and NetFlow record metadata. The raw data is often text (e.g., JSON, XML or human readable log formats) and must be parsed into a set of fields and values before being inserted into the database.


Designing an effective key scheme is vital to achieving efficient ingest and querying in Accumulo. Our storage schema is adapted from the Dynamic Distributed Dimensional Data Model (D4M) 2.0 schema~\cite{d4m2schema}. The LLCySA schema was originally described in~\cite{sawyer}. This section provides a brief summary and includes some extensions to the original design.
Entries in Accumulo are sorted by key, and data retrieval is performed through server-side scans of entries, coordinated by the Accumulo Java API. When a scan is initiated on a table, it is given a starting and ending
row ID range (e.g. ``a'' to ``b'') and will only return those entries
whose row IDs fall within that range. Column qualifiers can also be provided through the API to perform
projections on returned data. 
Because entries are sorted first by row ID, 
queries based primarily on row IDs are extremely fast. Therefore, row IDs must be designed such that the most common queries can be performed by simply selecting a range of rows.

At ingest time, performance can be impacted by hot spots in data distribution. Thus we employ a database sharding technique that uniformly and randomly distributes entries across all tablet servers. Sharding is achieved by prepending the row ID with a random zero-padded shard number between 0 and $N-1$, for a total of $N$ shards. The shard count is specified when the database is initialized, and experiments have indicated that $N$ should be at least as large as half the number of parallel client processes used for ingest.

\begin{figure}
\centering
\includegraphics[width=0.40\textwidth]{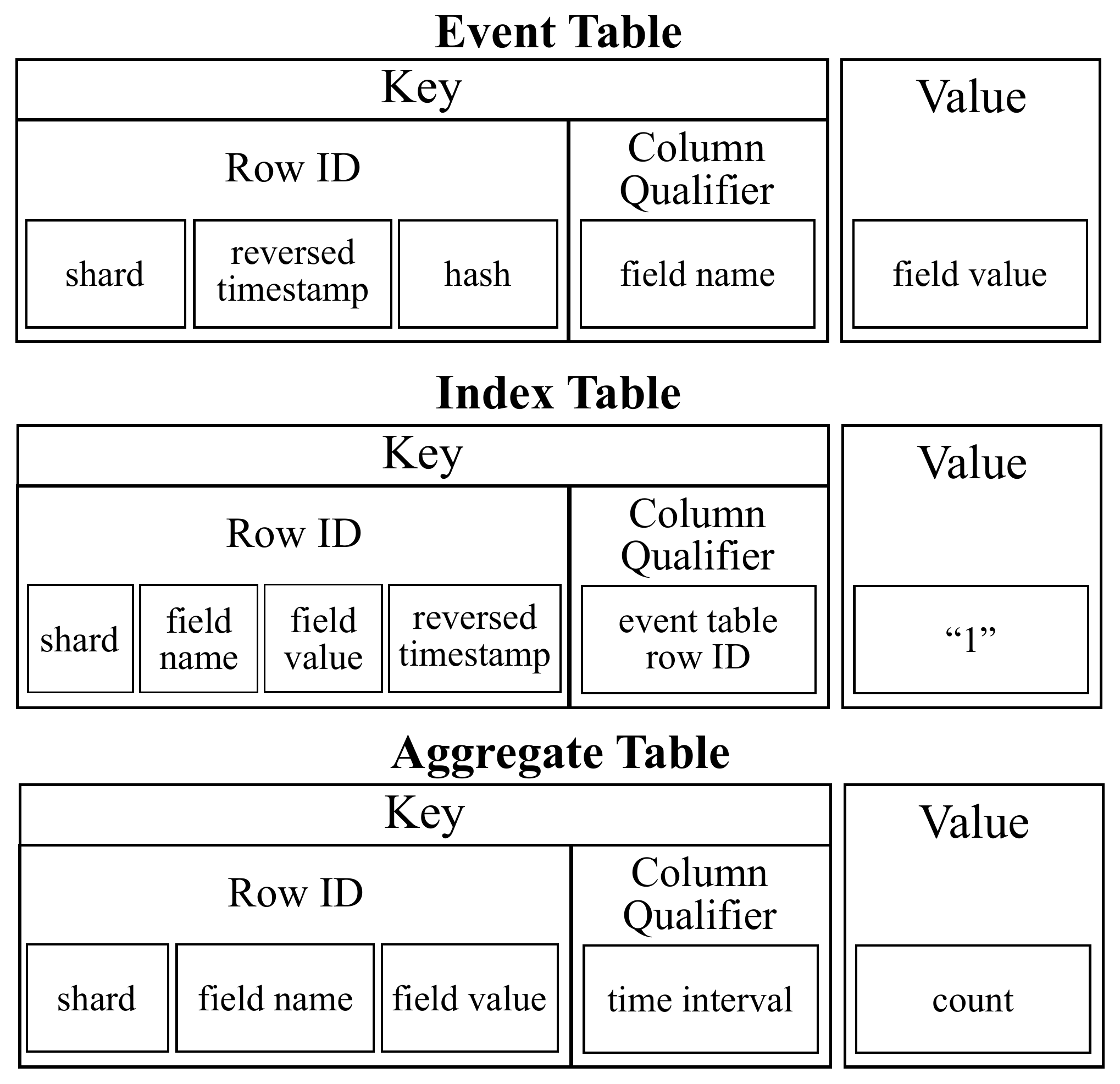}
\caption{The LLCySA storage schema uses three tables per data source with carefully designed keys to support highly parallel ingest and efficient retrieval.}
\label{fig:keyScheme}
\end{figure}

Our storage schema uses three tables per data source. The first table stores the primary event data. The key contains the shard number, a reversed timestamp to provide first-class support for filtering entries by time range, and a short hash to prevent collisions. The index table encodes field names and values in the row ID to allow fast look-ups by column value. The row ID stored in the index table's column qualifier matches the corresponding row in the event table. Finally, the aggregate table maintains a count of particular value occurrences by time interval. These counts are helpful in query planning. 
Figure~\ref{fig:keyScheme} shows the key scheme for each of the three table types.

Earlier stages in the big data pipeline collect and pre-process the various data sources, placing files to be ingested on a central filesystem. A master ingest process monitors new data and appends these files to a partitioned queue. Multiple ingest worker processes monitor a queue partition for work. Upon receiving a filename and metadata (i.e., data source and table name), the ingest worker reads lines from the file, parsing the data into entries to be stored in the event, index and aggregate tables. Counts for the aggregate table are summed locally by the ingest worker to reduce the number of records that must be aggregated on the server side using Accumulo's combiner framework. The master and worker processes are implemented in Python and use the Java Native Interface (JNI) to the Accumulo API. Entries are sent to Accumulo using the \texttt{BatchWriter} API class, which automatically batches and sends bulk updates to the database instance for efficiency.

Recall that our sharding scheme means that each entry can be routed to any of the tablet servers. 
This prevents hotspots based on timestamp locality during streaming ingest. Accumulo is designed for high parallelism, and many ingest worker processes are required to achieve high ingest rates. Our experiments in Sec.~\ref{sec:results} quantify the relationship between the number of client processes and ingest rates.

\section{Query Processing}
\label{sec:query}

In LLCySA, queries are specified by providing an event table, a time range, an optional list of columns to be retrieved, and an optional set of filtering conditions described by an arbitrary syntax tree. 
While our keying scheme supports this broad class of queries, there are still some challenges associated with data retrieval. First, scans over very large ranges of records can result in high memory utilization on the tablet servers and long latency in receiving initial query results. Second, joining the event and index tables for a particular data source can require handling a very large number of entries at the client. Our adaptive query batching and query planning techniques solve those two problems.

\subsection{Adaptive Query Batching}

Accumulo is designed for petabyte-scale data analysis and has been demonstrated to solve 
problems 
requiring 10--100 hours to complete~\cite{nsa}. Thus it is not tuned for low-latency performance on interactive queries. Scan results are sent to the client in batches of automatically determined size. This batching can impose a latency penalty of several seconds before the client receives the first result. Moreover, there is no capability similar to SQL's \texttt{LIMIT} statement in the Accumulo API. Therefore, without batched queries, the user may need wait for a large number of results to be buffered on the tablet servers even if only a small number of results are desired.

To improve interactive query performance, we have implemented an adaptive batching technique used on all queries based on the time range. 
Recall, all queries must specify a time range to restrict results. 
Instead of executing the entire query over the range $[t_{\text{start}},t_{\text{stop}}]$, the time range is partitioned into batches of smaller time ranges of size $b_0$, $b_1$, ..., etc., with each interval starting at position $p_0$, $p_1$, ..., etc., respectively.
Thus, the $i^{\text{th}}$ query batch restricts time to the interval 
$[p_i, p_i+b_i]$.
The size of each batch is selected to return approximately $k_i$ results. When the query batch is executed, it actually returns $r_i$ result rows and takes runtime $T_i$ to complete. These observed data points can then be used to adapt the batch size. The batch size is intended to be increased by a factor of $c$ each iteration, such that $k_{i+1}=ck_i$. However, the size will be adjusted if the next batch is expected to take too long or too short a time to run.
The batch size is adjusted when the estimated runtime $\hat{T}_{i+1}$ is not between $T_{\text{min}}$ and $T_{\text{max}}$, which are configurable settings. Alg.~\ref{alg:update} describes how the next batch is selected (where $\epsilon$ refers to the minimum time resolution).


\begin{algorithm}
\caption{Update Batch Parameters}
\label{alg:update}

\renewcommand{\algorithmicrequire}{\textbf{Input:}}
\renewcommand{\algorithmicensure}{\textbf{Output:}}

\begin{algorithmic}[1]

 \Require Runtime in seconds of the last query batch ($T_i$) and number of results returned from last query batch ($r_i$)
 \Ensure Parameters for batch $i+1$ are set.

\Function{Update}{ $ T_i, r_i $ }

 \State $k_{i+1} \leftarrow ck_i$ \Comment{desired result count for next batch}
 
 \State $\hat{T}_{i+1} \leftarrow k_{i+1}(T_i/r_i)$ \Comment{estimate batch runtime}

 \If{$ \hat{T}_{i+1} > T_{\text{max}} $} 
  \State $k_{i+1} \leftarrow T_{\text{max}} ( r_i / T_i ) $ \Comment{batch is too large}
 \ElsIf{$ \hat{T}_{i+1} < T_{\text{min}} $} 
  \State $k_{i+1} \leftarrow T_{\text{min}} ( r_i / T_i ) $ \Comment{batch is too small}
 \EndIf


 \State $ b_{i+1} \leftarrow \text{min} ( k_{i+1}(b_i/r_i) , t_{\text{stop}}-p_i ) $ \Comment{set batch size}

 \State $ p_{i+1} \leftarrow p_i + b_i + \epsilon $ \Comment{update position}

\EndFunction
\end{algorithmic}

\end{algorithm}

The query is initialized with $k_0=10$, $p_0=t_{\text{start}}$, and $b_0$ pre-computed for the particular Accumulo table being queried based on the typical hit-rates $r_i/b_i$ of previous queries on that table.
While this initial estimate $b_0$ may be inaccurate, subsequent batch sizes adjust based on each partial query's runtime.
By default, $c=1.5$, $T_{\text{max}}=30\,\text{s}$, and $T_{\text{min}}=1\,\text{s}$. Batches are executed according to Alg.~\ref{alg:batch}, where the function \texttt{query} takes a time range as input and returns the runtime of the query and the number of rows in the result set. 

\begin{algorithm}
\caption{Batched Query}
\label{alg:batch}
\begin{algorithmic}[1]

 \While{$p_i < t_{\text{stop}} $}

 	\State $ (T_i, r_i) \leftarrow \text{query}(p_i, p_i+b_i) $
 	\State $\text{update}(T_i, r_i)$

 	\State $i \leftarrow i+1$
 \EndWhile

\end{algorithmic}

\end{algorithm}

In addition to returning the first result rows more quickly, query batching has the benefit of ensuring the most recent results are generally (but not strictly) returned to the user first. Due to sharding, all queries utilize the Accumulo API's \texttt{BatchScanner}, which makes no guarantee on the ordering of results. Essentially, results are returned from each~tablet server as they become available. Without time range batching, uneven load on the tablet servers could result in highly unordered result sets.

\subsection{Query Planning}

Query planning is the process of selecting an efficient set of steps for executing a specified query. In a RDBMS, queries are specified using SQL, a declarative language with which users describe their desired result set rather than a procedure for obtaining those results. Query planning and optimization in this context, particularly for SQL queries involving joins between multiple tables, is challenging and has been a focus of research for more than 30 years~\cite{chaudhuri1998overview}.

Fortunately, query planning in typical Accumulo applications is simple because the set of supported query operations is far less expressive than SQL. In particular, queries are restricted to a single data source. In the LLCySA query processor, queries are specified by projecting onto a subset of columns and by restricting rows based on an arbitrary set of conditions. These conditions take the form of a syntax tree, where each node is a boolean operation (i.e., ``and'', ``or'', and ``not'') or a conditional statement applied to a particular field-value pair. Conditions can enforce equality (e.g., \texttt{field1 = value1}), inequality (e.g., \texttt{field1 < value1}) or regular expression matching (e.g., \texttt{field1} matches \texttt{regex1}). Query planning in this context is more accurately described as access path selection, and a naive but effective solution only requires a cost model for scans of the index and event tables~\cite{selinger1979access}.

The index table can be used to efficiently evaluate equality conditions for indexed fields. Therefore, the LLCySA query planner selects certain equality condition nodes from the filter syntax tree and executes those conditions via index table scans. The remaining syntax tree is executed by filtering on the tablet servers, using a custom subclass of Accumulo's \texttt{WholeRowIterator}. The query planner uses the aggregate table for selectivity estimation, by assigning each equality condition a factor $d$, which is a density estimate related to the inverse of selectivity. A global, empirically derived parameter, $w$, determines a threshold to avoid intersections between sets of significantly different sizes.

LLCySA plans a query described by filter syntax tree $T$ with conditions $c_i \in T$, where each equality condition is assigned a density $d_i$. 
A set of simple heuristics are used to select equality conditions for index scans. The key sets from index scans are combined via intersection or union, and all other conditions are evaluated using tablet server filtering. The resulting planner is capable of efficiently processing the queries described by LLCySA use cases. The heuristics are defined as follows:
\begin{itemize}
\item If the tree root is an equality condition, then use an index scan.
\item If the tree root is a logical ``or'' and all children are equality conditions, then use an index scan for all conditions and union the resulting key sets.
\item If the tree root is a logical ``and'', use an index scan for all children whose density is less than $w\min_{i} d_i$ (for all $i$ in the set of equality conditions whose parent is the root node), intersect the resulting key sets, and pass those results to an event scanner.
\item Otherwise, evaluate the entire syntax tree using tablet server filtering.
\end{itemize}

\begin{figure}
\centering
\includegraphics[width=0.5\textwidth]{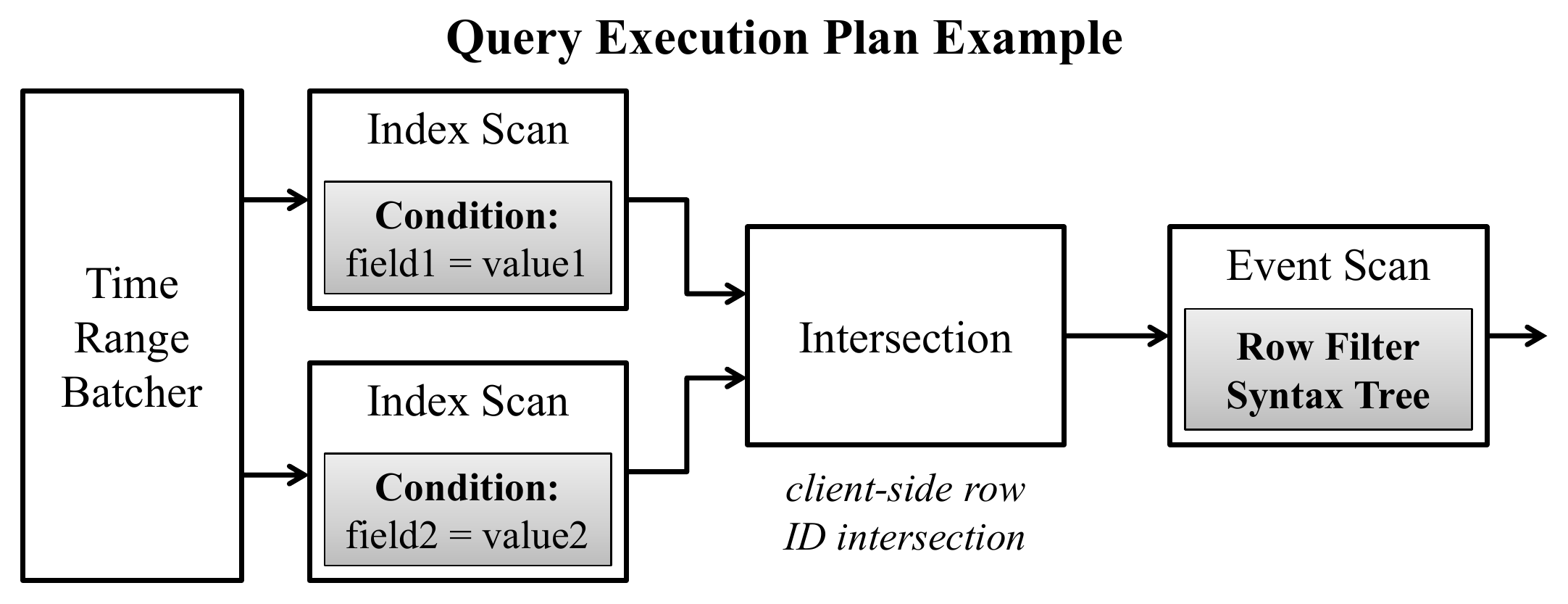}
\caption{In this notional query execution plan, the query is restricted by multiple conjoined (i.e., intersected) conditions. Equality conditions can be evaluated using the index table with row IDs intersected at the client. Resulting row IDs can then be passed to an event table scanner, where additional filter conditions can be applied.}
\label{fig:planning}
\end{figure}

Fig.~\ref{fig:planning} shows a notional query execution plan in which two conjoined equality conditions are evaluated using the index table and the remaining conditions are evaluated using tablet server filters. This simple approach avoids the need for the far richer cost models, query re-writing techniques, and robust selectivity estimation used in a modern RDBMS~\cite{hellerstein2005readings}. Note, the entire query execution is batched according to Algs.~\ref{alg:update} and~\ref{alg:batch}.

\section{Results}
\label{sec:results}

The LLCySA research system runs on a 24-node cluster of commodity servers, each containing 24 cores, 64 GB of memory, and a 12 TB hard disk RAID. 
Data is staged on separate storage running the Lustre parallel filesystem.
The cluster is interconnected on a 10 Gbps Ethernet switch. 
Query experiements were conducted on a single 8-node Accumulo 1.4.1 instance.
For ingest experiments, Accumulo 1.5.0 database instances with various numbers of tablet servers were deployed using a database provisioning system. Ingest processes are dispatched by a scheduler.

The query experiments in this paper focus on web traffic captured from web proxy server log files. Each event occurrence represents a single HTTP request and has dozens of attributes, including timestamp, source and destination IP addresses, 
requested domain name, 
and URL. This is an interesting experimental data source due to its high dimensionality and volume, and its potential utility for analysis.

\subsection{Ingest Scaling}

\begin{figure}
\centering
\includegraphics[width=0.5\textwidth]{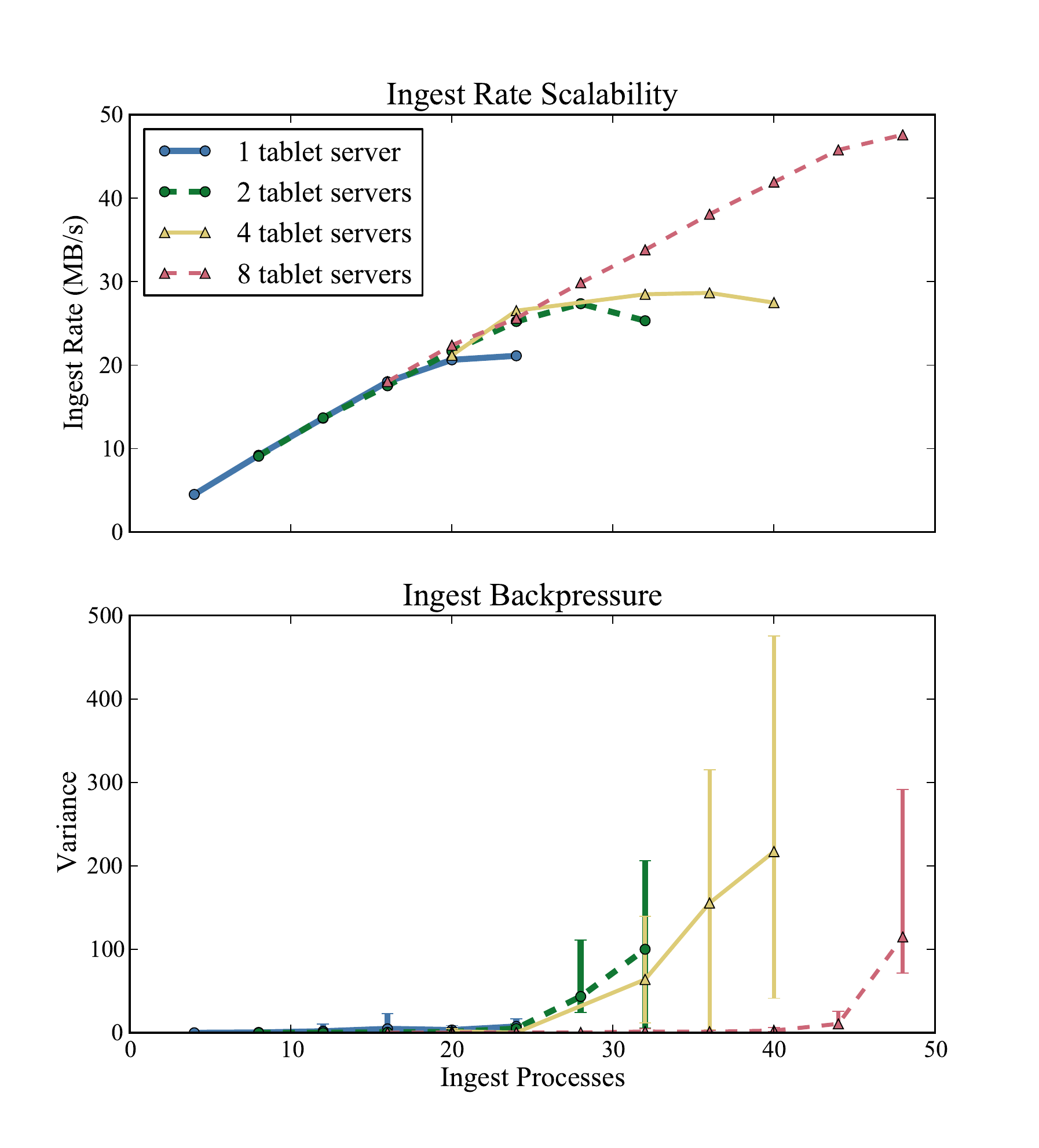}
\caption{Ingest rates scale linearly with the number of client processes sending entries to the tablet servers. Ingest rates level off depending on the number of tablet servers. As tablet servers saturate, the variance of the ingest rates (vs. time) increase significantly.}
\label{fig:scaling}
\end{figure}

In this experiment, we vary the number of the processes sending entries to Accumulo as well as the number of tablet servers in the Accumulo instance. 
Results show that ingest rates scale linearly with the number of client processes, as shown in Fig.~\ref{fig:scaling} (top). The saturation point for ingest rates varies based on the number of tablet servers in the Accumulo instance. For sustained ingest, the number of tablet and servers client processes should be selected to keep the database instance operating well below this saturation point. As ingest rates approach the saturation point, tablet servers apply backpressure to ingestors (Fig.~\ref{fig:scaling}, bottom). Backpressure is measured as the variance of the steady-state time series ingest rates.

\begin{figure*}
\centering
\begin{subfigure}[]
	\centering
	\includegraphics[width=0.32\textwidth]{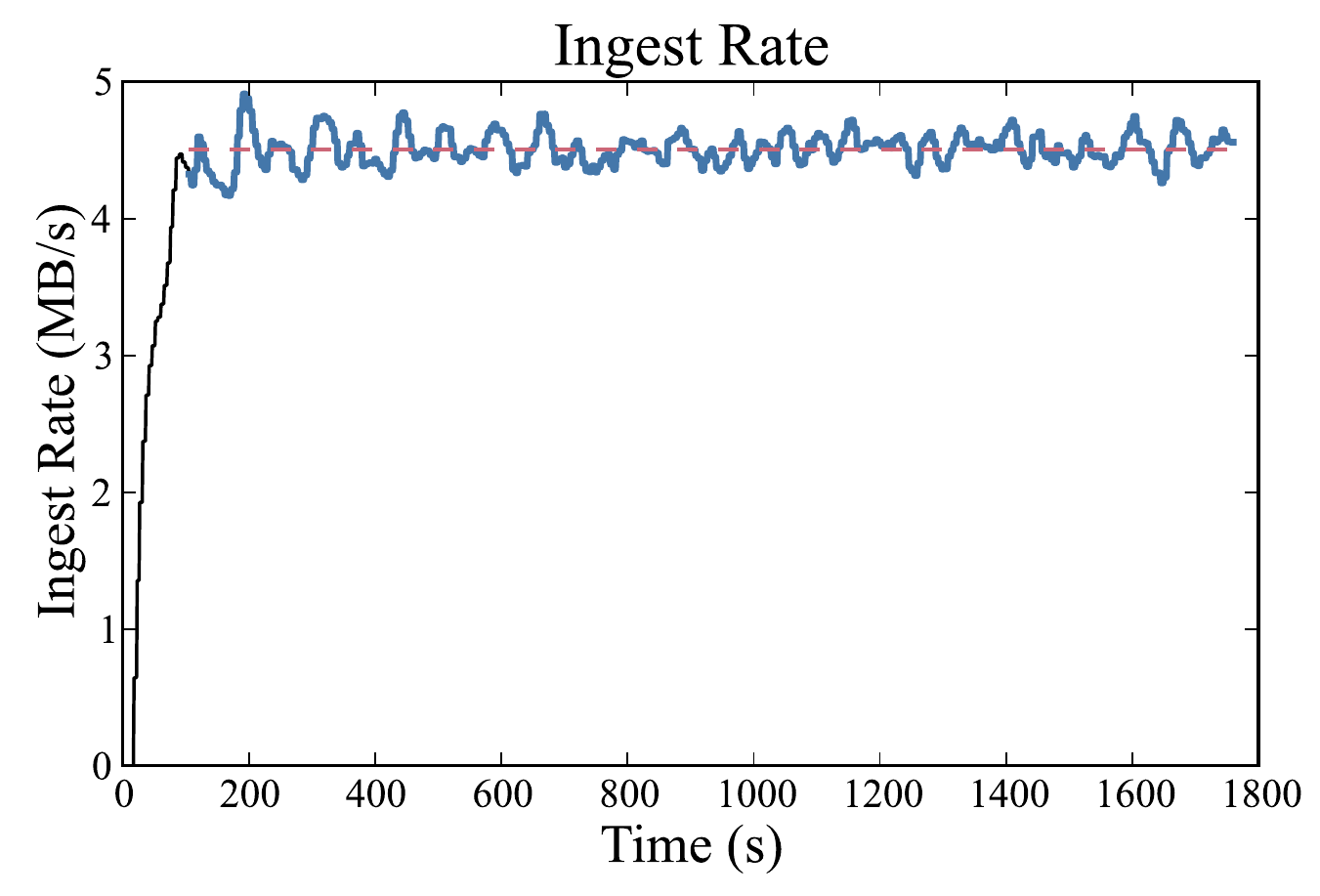}
\end{subfigure}
\begin{subfigure}[]
	\centering
	\includegraphics[width=0.32\textwidth]{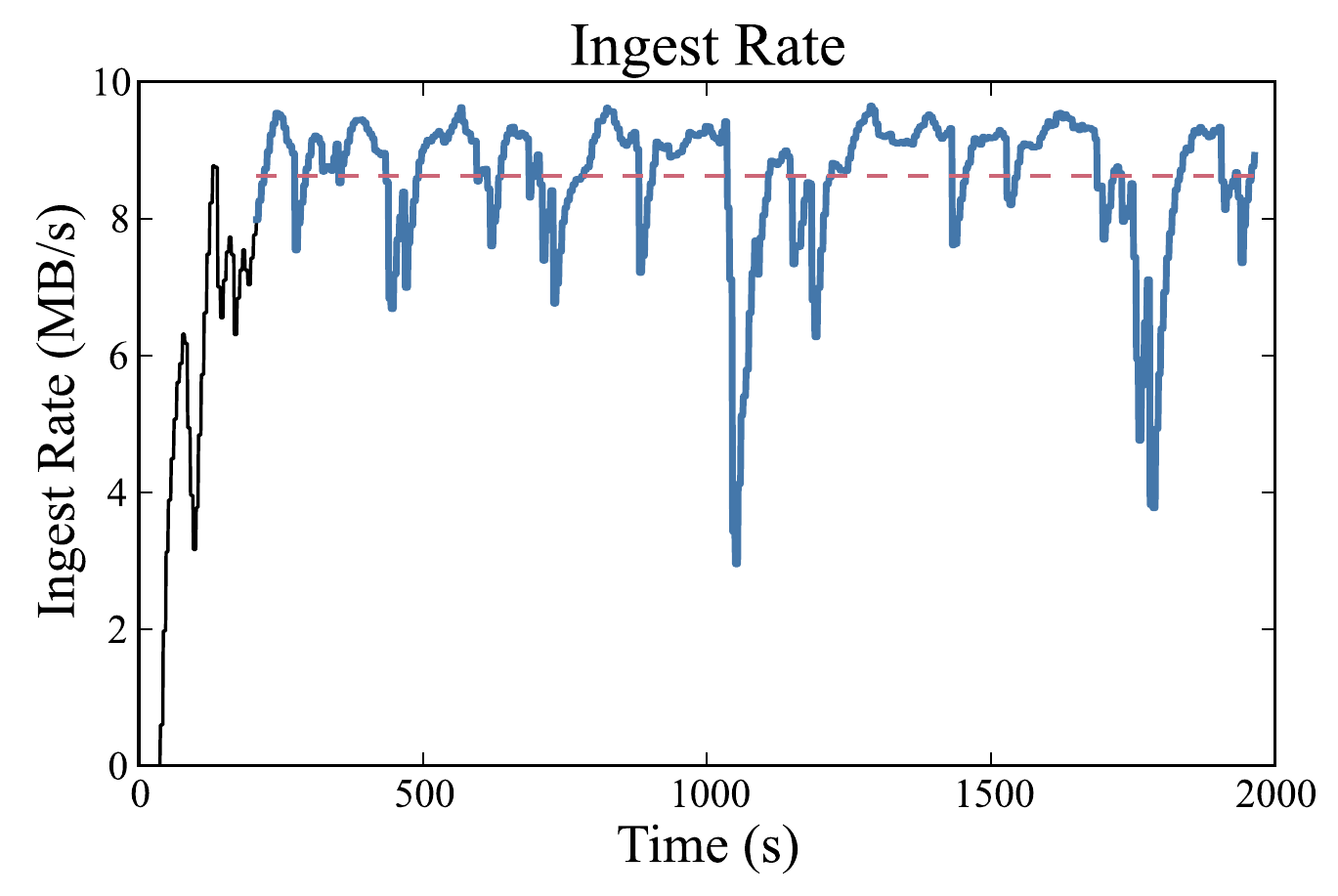}
\end{subfigure}
\begin{subfigure}[]
	\centering
	\includegraphics[width=0.32\textwidth]{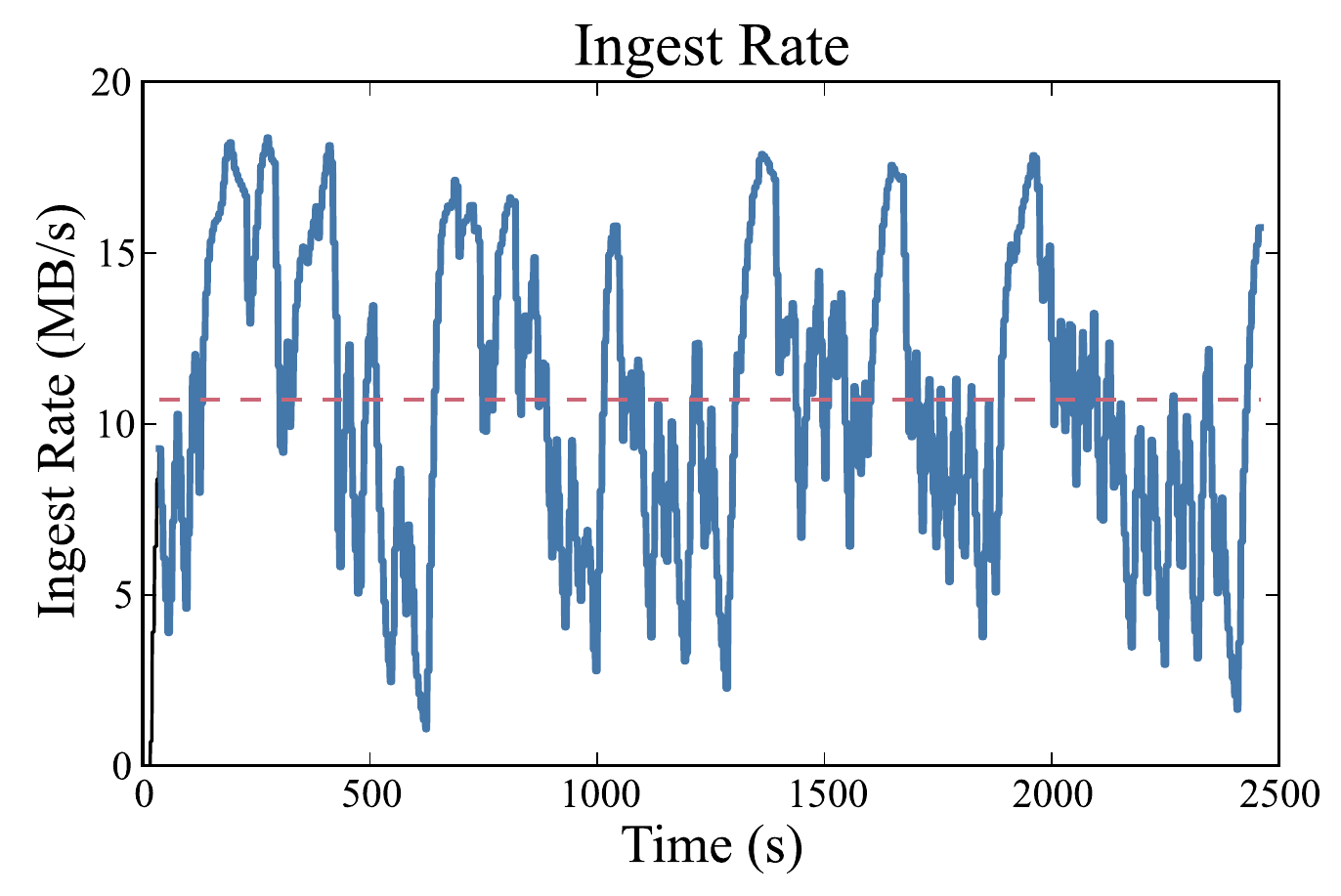}
\end{subfigure}
\caption{Time series ingest rates are an indicator of database conditions. Plots show steady-state ingest rates (blue line) as well as steady-state mean ingest rates (red dashed line). At low ingest workloads (a), rates maintain a level with low variance. As rates approach the limit for the database instance (b), backpressure can be observed as dips in the rate plot. Finally, when tablet servers are saturated (c), ingest rates exhibit high variation.}
\label{fig:timeseries}
\end{figure*}

Instantaneous ingest rates with respect to time are good indicators of the state of the tablet servers. When ingest rates are below capacity of the database instance, rates have a relatively smooth time series. As ingest rates approach saturation, the tablet servers create backpressure by blocking ingest processes while memory-cached entries must be written to disk. When ingest rates are at or above capacity, ingest rates exhibit high variation, and the database instance can typically not sustain these rates. Fig.~\ref{fig:timeseries} shows time series plots for each of these cases.

\subsection{Query Responsiveness}

In order to increase query responsiveness, scan ranges are batched into sub-ranges of increasing size. The size of batches are determined adaptively based on the runtime of the last partial query and the number of results returned. This experiment measures the responsiveness of a set of queries that select web traffic to a particular domain name. Queries A, B and C select traffic to the most popular domain name, a somewhat popular domain name, and an unpopular domain name, respectively. Each query is also restricted by time range. Responsiveness is measured based on the amount of latency before the user receives the first entry, 100th entry, and 1000th entry. The runtime of the total query is also measured. 

Each query is executed using four different schemes. First, a simple scan (``Scan'') is initiated for the full time range, and tablet server filtering is used to select rows in the web traffic table that match the requested domain condition (rather than using the index table). Next, this scan is batched (``Batched Scan'') using our adaptive technique, but rows are still filtered by the scan rather than using the index table. In the third scheme (``Index''), query planning is enabled and the index table is used to find all traffic to the requested domain name during the query time range. Then the corresponding rows are retrieved from the main table. In this scheme, we do not batch the scans on either the index or event tables. In the final scheme, query planning is used along with the adaptive batching technique (``Batched Index'').

\begin{figure*}
\centering
\begin{subfigure}[]
	\centering
	\includegraphics[width=0.32\textwidth]{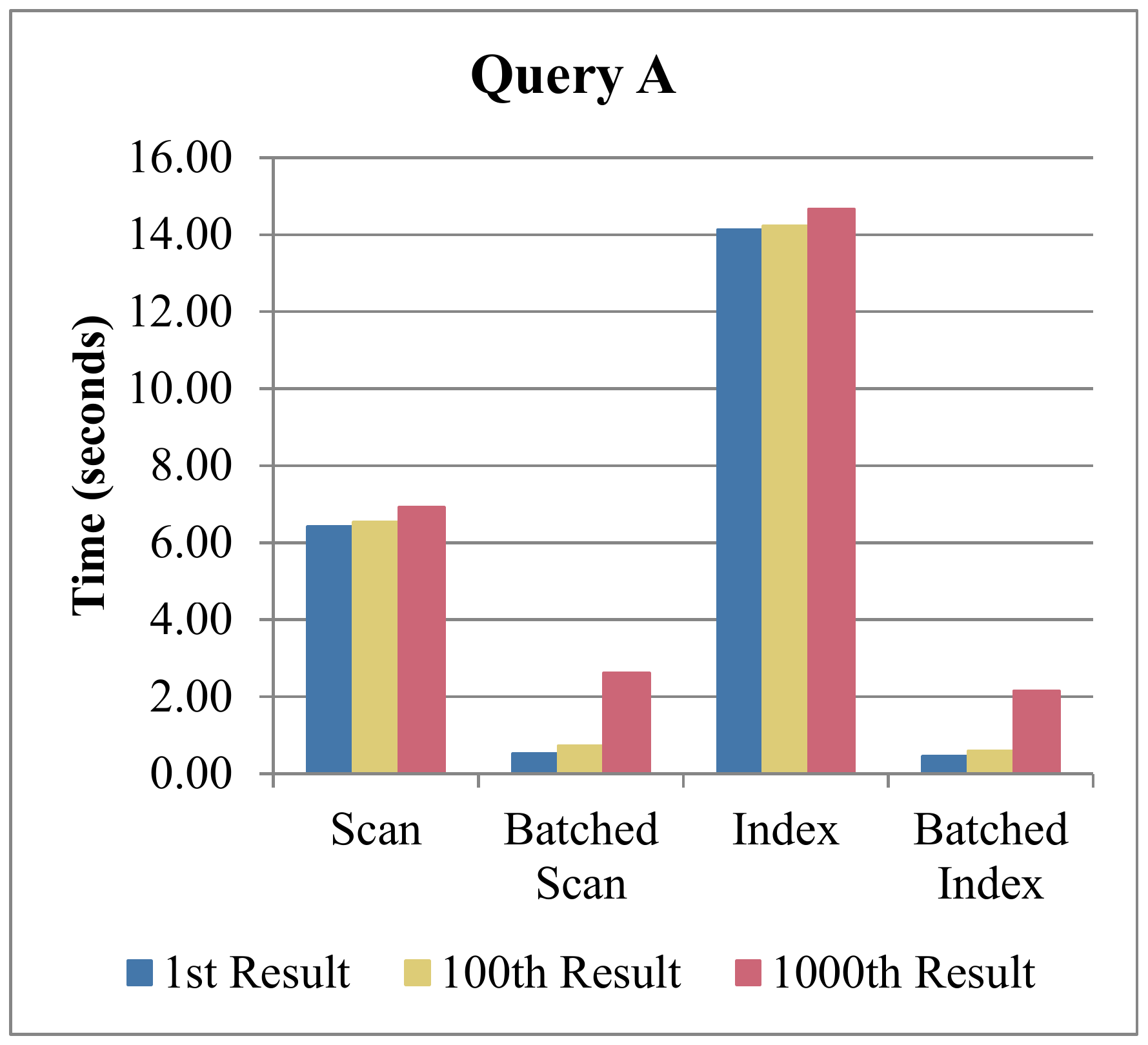}
\end{subfigure}
\begin{subfigure}[]
	\centering
	\includegraphics[width=0.32\textwidth]{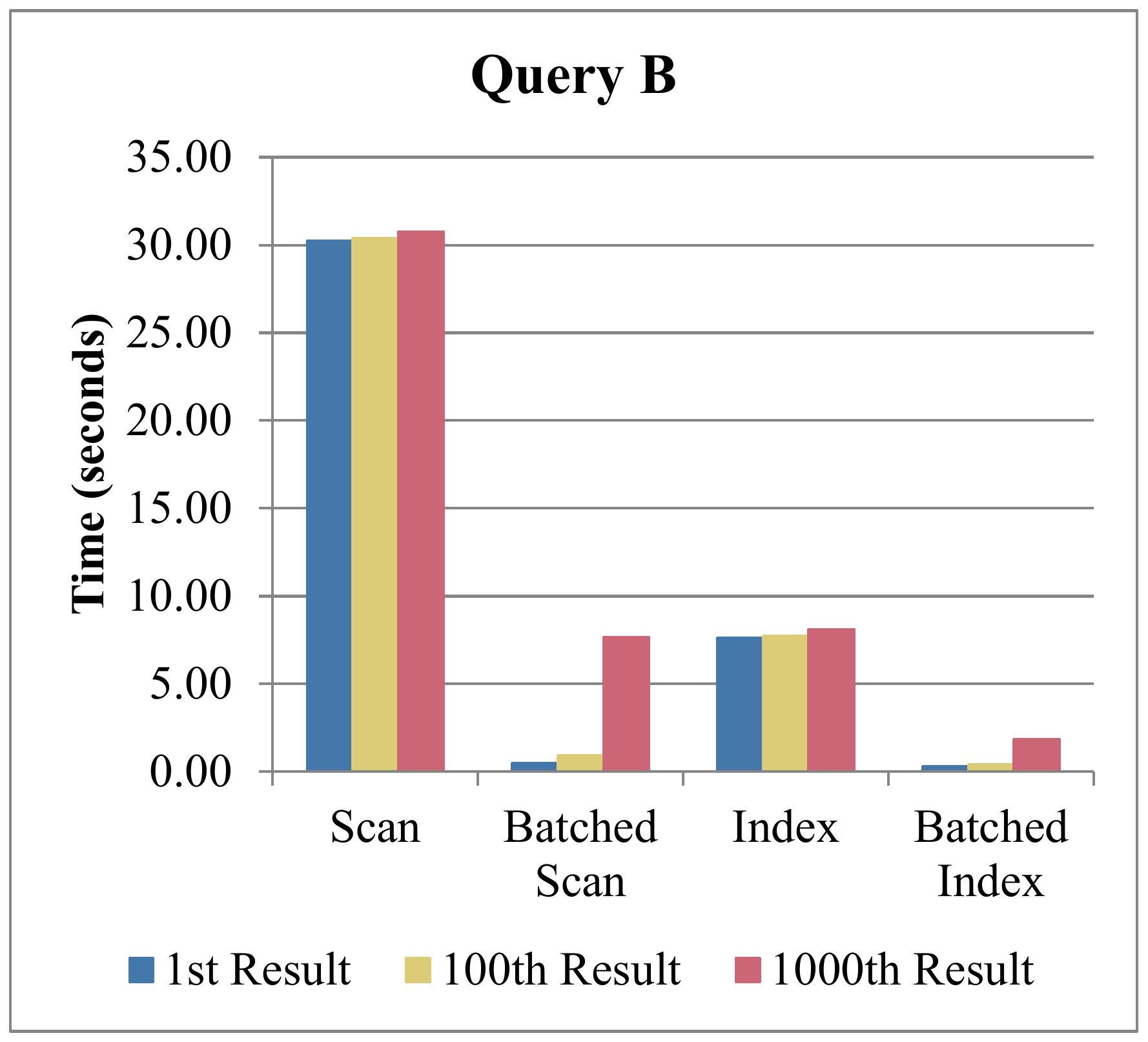}
\end{subfigure}
\begin{subfigure}[]
	\centering
	\includegraphics[width=0.32\textwidth]{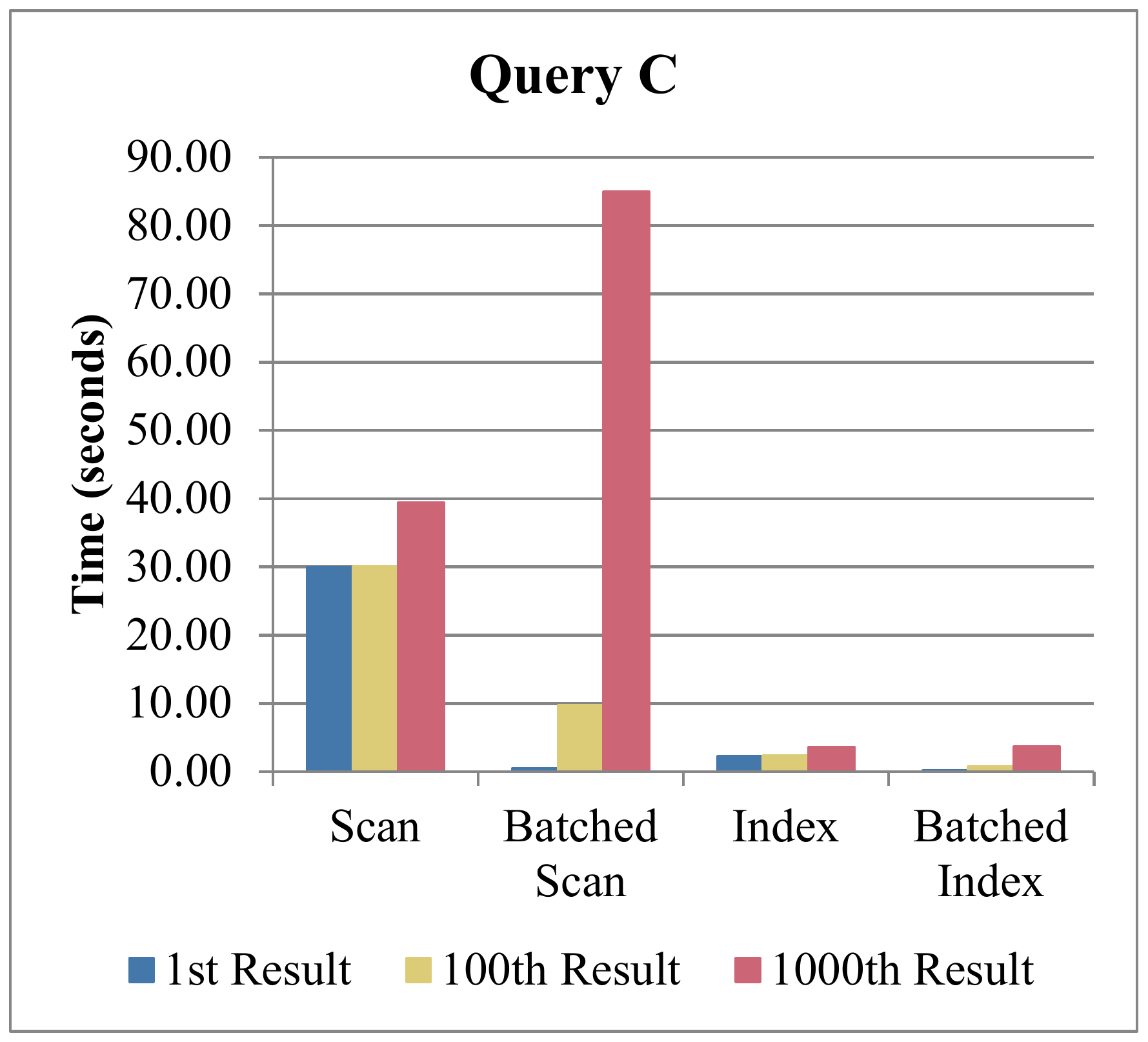}
\end{subfigure}
\caption{Three queries are executed using four different strategies. For each query, adaptive batching using the index table (``batched index'') was most responsive, delivering the first result with minimum latency.}
\label{fig:adaptive}
\end{figure*}

Fig.~\ref{fig:adaptive} shows the responsiveness of the three test queries using each of the four query execution techniques. In Query A, 
the batched scan and batched index technique achieve roughly equivalent performance because the requested domain in this query is extremely common. Generally, indexing offers little benefit in finding values that occur very frequently. When combined with other filter conditions, the query planner typically does not utilize the index table for extremely common values. In Query B, 
both batched scan and batched index return initial results with low latency, but for larger result sets, indexing improves performance. Finally, in Query C, 
index and batched index deliver the best performance because very few rows match the query conditions. Table~\ref{tab:responsive} shows the responsiveness for all queries schemes, measured as the latency for the first result. For all three queries, batched indexing provided the fastest first result.

\begin{table}
\small
\caption{Query Responsiveness (time to first result in seconds)}
\label{tab:responsive}
\begin{center}
  \begin{tabular}{ | l || c | c | c | }
    \hline
    & Query A & Query B & Query C \\ \hline
    Scan          & 6.42  & 30.25 & 29.91  \\ \hline
    Batched Scan  & 0.52  & 0.47 & 0.49 \\ \hline
    Index         & 14.13 & 7.62 & 2.24 \\ \hline
    \textbf{Batched Index} & \textbf{0.45}  & \textbf{0.27} & \textbf{0.16} \\
    \hline
  \end{tabular}
\end{center}
\end{table}

\begin{table}
\small
\caption{Query Runtime to Completion (seconds)}
\label{tab:query}
\begin{center}
  \begin{tabular}{ | l || c | c | c | }
    \hline
    & Query A & Query B & Query C \\ \hline
    Scan & 316.4 & 291.3 & 286.2 \\ \hline
    Batched Scan & 312.6 & 294.8 & 298.0 \\ \hline
    Index & 94.5 & 29.4 & 6.7 \\ \hline
    Batched Index & 96.8 & 31.7 & 5.9 \\
    \hline
  \end{tabular}
\end{center}
\end{table}

Therefore, adaptive batching and indexing achieves the most responsive query results. However, batching introduces some overhead, which can result in the entire query taking longer to complete. Table~\ref{tab:query} shows the full runtime for each query, executed for a time range of four hours of web traffic. While batching increases overall runtime in most cases, the overhead can be considered negligible for interactive applications.

\section{Conclusions}
\label{sec:conclusions}

We present a data pipeline design for delivering real-time ingest and responsive queries using Apache Accumulo. Ingest scalability tests demonstrate the saturation point of various Accumulo database configurations and indicate that many parallel ingest clients are required to achieve high ingest rates. Additionally, we show that batching queries effectively lowers latency to the first result arriving at the client. 

Scalability is the primary design driver for Accumulo and NoSQL databases in general. While Accumulo has been shown to scale beyond 1,000 nodes, it is also an effective tool for smaller clusters. Indeed, the key benefit to Accumulo for LLCySA is ingest rate. This particular system can handle the velocity of the current set of data sources, and the system architecture can scale to accommodate larger networks or denser data sources by simply adding tablet servers and compute nodes. While a traditional RDBMS would offer advantages in terms of query performance and flexibility, Accumulo's insertion scalability is unmatched. 

For our particular configuration, ingest was shown to scale at a rate of 1.1 MB/s per client ingest process. This indicates a bottleneck in the client (i.e., the Accumulo API's \texttt{BatchWriter}). The scaling was observed to be quite linear until ingest rates approach a maximum supported load. This ingest limit is determined by the number of tablet servers in the Accumulo instance. Sustained ingest load should be restricted to avoid saturation, as this can ultimately degrade performance and stability.

NoSQL databases and technologies do not enforce a strict table schema, and this relaxed constraint is often touted as a key benefit. However, in a sorted key--value store, key selection impacts performance perhaps more than any other design decision. The LLCySA row-keying scheme effectively shards the data across the database instance and reverse-sorts events within a shard by timestamp. This design mitigates hotspots during data ingest while enabling queries to restrict by timestamp (a key feature for our situational awareness application) with essentially zero cost. Reasonable query performance also requires indexing and planning, standard features of RDBMSs that need to be re-implemented in Accumulo applications.

Achieving responsive query processing also requires batching queries into smaller sub-queries. Accumulo is tuned for long-running tasks, as even simple analytics on petabyte-scale datasets can take tens to hundreds of hours. When processing at this scale, overhead on the order of seconds is tolerable. However, for interactive queries, adaptively batching queries based on runtime and result set size gives the fine-grained control necessary for the desired responsiveness.

\section{Acknowledgments}
The authors would like to think the many people whose hard work helped make these experiments possible. The LLCySA ingest and query pipeline builds upon the work of our fellow members of the LLCySA team: 
Ethan Aubin, Suresh Damodaran, Chris Degni, Jack Fleischman, Jeff Gottschalk, Stephen Kelley, Jeremy Kepner, Jeremy Mineweaser, Matt Schmidt, Alexia Schulz, Diane Staheli, An Tran, Jim Will, and Tamara Yu. 
Our computing environment was developed and supported by the entire LLGrid team: 
William Arcand, David Bestor, Bill Bergeron, Chansup Byun, Matthew Hubbell, Peter  Michaleas, Julie Mullen, Andrew Prout, Albert Reuther, Antonio Rosa, and Charles Yee.

\bibliographystyle{IEEEtran}
\bibliography{IEEEabrv,bib/references}

\begin{thebibliography}{10}
\providecommand{\url}[1]{#1}
\csname url@samestyle\endcsname
\providecommand{\newblock}{\relax}
\providecommand{\bibinfo}[2]{#2}
\providecommand{\BIBentrySTDinterwordspacing}{\spaceskip=0pt\relax}
\providecommand{\BIBentryALTinterwordstretchfactor}{4}
\providecommand{\BIBentryALTinterwordspacing}{\spaceskip=\fontdimen2\font plus
\BIBentryALTinterwordstretchfactor\fontdimen3\font minus
  \fontdimen4\font\relax}
\providecommand{\BIBforeignlanguage}[2]{{%
\expandafter\ifx\csname l@#1\endcsname\relax
\typeout{** WARNING: IEEEtran.bst: No hyphenation pattern has been}%
\typeout{** loaded for the language `#1'. Using the pattern for}%
\typeout{** the default language instead.}%
\else
\language=\csname l@#1\endcsname
\fi
#2}}
\providecommand{\BIBdecl}{\relax}
\BIBdecl

\bibitem{accumulo}
\BIBentryALTinterwordspacing
\emph{Apache Accumulo}. [Online]. Available: \url{http://accumulo.apache.org}
\BIBentrySTDinterwordspacing

\bibitem{fuchslecture}
A.~Fuchs, ``Accumulo: extensions to {G}oogle's {B}igtable design,'' March 2012,
  lecture, Morgan State University.

\bibitem{patil2011ycsb}
S.~Patil, M.~Polte, K.~Ren, W.~Tantisiriroj, L.~Xiao, J.~L{\'o}pez, G.~Gibson,
  A.~Fuchs, and B.~Rinaldi, ``{YCSB}++: benchmarking and performance debugging
  advanced features in scalable table stores,'' in \emph{Proc. of the 2nd ACM
  Symposium on Cloud Computing}.\hskip 1em plus 0.5em minus 0.4em\relax ACM,
  2011, p.~9.

\bibitem{llcysa2014}
S.~Sawyer, T.~Yu, B.~O'Gwynn, and M.~Hubbell, ``{LLCySA}: Making sense of
  cyberspace,'' \emph{Lincoln Laboratory Journal}, vol.~20, no.~2, 2014.

\bibitem{nsa}
P.~Burkhardt and C.~Waring, ``An {NSA} {B}ig {G}raph experiment,'' National
  Security Agency, Tech. Rep. NSA-RD-2013-056002v1, 2013.

\bibitem{sen2013benchmarking}
R.~Sen, A.~Farris, and P.~Guerra, ``Benchmarking apache accumulo bigdata
  distributed table store using its continuous test suite,'' in \emph{IEEE
  International Congress on Big Data}, 2013, pp. 334--341.

\bibitem{husain2011heuristics}
M.~Husain, J.~McGlothlin, M.~M. Masud, L.~Khan, and B.~Thuraisingham,
  ``Heuristics-based query processing for large {RDF} graphs using cloud
  computing,'' \emph{Knowledge and Data Engineering, IEEE Transactions on},
  vol.~23, no.~9, pp. 1312--1327, 2011.

\bibitem{abouzeid2009hadoopdb}
A.~Abouzeid, K.~Bajda-Pawlikowski, D.~Abadi, A.~Silberschatz, and A.~Rasin,
  ``{HadoopDB}: an architectural hybrid of {MapReduce} and {DBMS} technologies
  for analytical workloads,'' \emph{Proceedings of the VLDB Endowment}, vol.~2,
  no.~1, pp. 922--933, 2009.

\bibitem{storm}
\BIBentryALTinterwordspacing
\emph{Apache Storm}. [Online]. Available:
  \url{http://storm.incubator.apache.org/}
\BIBentrySTDinterwordspacing

\bibitem{kepner2012dynamic}
J.~Kepner, W.~Arcand, W.~Bergeron, N.~Bliss, R.~Bond, C.~Byun \emph{et~al.},
  ``Dynamic distributed dimensional data model ({D4M}) database and computation
  system,'' in \emph{Proc. of IEEE Conf. on Acoustics, Speech and Signal
  Processing (ICASSP)}, 2012.

\bibitem{sandia}
B.~Wylie, D.~Dunlavy, W.~Davis, and J.~Baumes, ``Using {NoSQL} databases for
  streaming network analysis,'' in \emph{Large Data Analysis and Visualization
  (LDAV)}.\hskip 1em plus 0.5em minus 0.4em\relax IEEE, 2012, pp. 121--124.

\bibitem{d4m2schema}
J.~Kepner, C.~Anderson, W.~Arcand, D.~Bestor, B.~Bergeron, C.~Byun
  \emph{et~al.}, ``{D4M} 2.0 schema: A general purpose high performance schema
  for the {A}ccumulo database,'' in \emph{Proc. of High Performance Extreme
  Computing (HPEC)}.\hskip 1em plus 0.5em minus 0.4em\relax IEEE, 2013.

\bibitem{sawyer}
S.~Sawyer, D.~O'Gwynn, A.~Tran, and T.~Yu, ``Understanding query performance in
  {Accumulo},'' in \emph{Proc. of High Performance Extreme Computing
  (HPEC)}.\hskip 1em plus 0.5em minus 0.4em\relax IEEE, 2013.

\bibitem{chaudhuri1998overview}
S.~Chaudhuri, ``An overview of query optimization in relational systems,'' in
  \emph{Proceedings of ACM SIGACT-SIGMOD-SIGART symposium on Principles of
  Database Systems}, 1998, pp. 34--43.

\bibitem{selinger1979access}
P.~G. Selinger, M.~M. Astrahan, D.~D. Chamberlin, R.~A. Lorie, and T.~G. Price,
  ``Access path selection in a relational database management system,'' in
  \emph{Proceedings of ACM SIGMOD International Conference on Management of
  Data}, 1979, pp. 23--34.

\bibitem{hellerstein2005readings}
J.~M. Hellerstein and M.~Stonebraker, \emph{Readings in database
  systems}.\hskip 1em plus 0.5em minus 0.4em\relax MIT Press, 2005.

\end{thebibliography}
%



\end{document}